# Effects of Cr substitution on the magnetic and transport properties and electronic states of SrRuO$_3$ epitaxial thin films

Kaori Kurita[1], Akira Chikamatsu[1,2,*], Kei Shigematsu[1,3], Tsukasa Katayama[1], Hiroshi Kumigashira[4], Tomoteru Fukumura[2,5], and Tetsuya Hasegawa[1,2,3]

[1]*Department of Chemistry, The University of Tokyo, Tokyo 113-0033, Japan*

[2]*CREST, Japan Science and Technology Agency (JST), Tokyo 113-0033, Japan*

[3]*Kanagawa Academy of Science and Technology (KAST), Kawasaki 213-0012, Japan*

[4]*Institute of Materials Structure Science, KEK, Tsukuba 305-0801, Japan*

[5]*Department of Chemistry, Tohoku University, Sendai, Miyagi 980-8578, Japan*

(Dated: September 9, 2015)

## Abstract

The effect of Cr substitution in a SrRuO$_3$ epitaxial thin film on SrTiO$_3$ substrate was investigated by measuring the magnetic and transport properties and the electronic states. The ferromagnetic transition temperature of the SrRu$_{0.9}$Cr$_{0.1}$O$_3$ film (166 K) was higher than that of the SrRuO$_3$ film (147 K). Resonant photoemission spectroscopy experimentally revealed that the Cr 3$dt_{2g}$ orbital is hybridized with the Ru 4$dt_{2g}$ orbital in the SrRu$_{0.9}$Cr$_{0.1}$O$_3$ film, supporting the assumption that the enhancement of the ferromagnetic transition temperature through Cr substitution stems from the widening of energy bands due to the hybridization of Cr 3$dt_{2g}$ and Ru 4$dt_{2g}$ orbitals. Furthermore, we found that the Hall resistivity of the SrRu$_{0.9}$Cr$_{0.1}$O$_3$ film at low temperature is not a linear function of magnetic field in the high-field region where the out-of-plane magnetization was saturated; this result suggests that the SrRu$_{0.9}$Cr$_{0.1}$O$_3$ film undergoes a structural transition at low temperature accompanied with the modulation of the Fermi surface.

---

[*] Corresponding author. E-mail: chikamatsu@chem.s.u-tokyo.ac.jp

# I. INTRODUCTION

Perovskite ruthenium oxide, SrRuO$_3$ (SRO), which is a ferromagnetic metal with a Curie temperature $T_C$ = 163 K, has been extensively studied for its potential application as an electrode material in ferroelectric and magnetoresistive random access memory devices.[1-3] The itinerant ferromagnetism of SRO originates from a narrow Ru $4dt_{2g}$ band, which can be drastically modified through local distortion by substituting other elements for the Sr site.[4] For example, Ca substitution suppresses ferromagnetism in Ca$_{1-x}$Sr$_x$RuO$_3$, and $T_C$ decreases to zero at $x$ = 0.4.[5] Photoemission spectroscopy has revealed that with Ca substitution, a spectral weight is transferred from the coherent part to the incoherent part as a consequence of increased correlation strength due to the lifting of the $t_{2g}$ orbital degeneracy without changing the bandwidth of Ru $4d$.[6] The Ru sites of SRO have also been substituted with various transition metal ions such as Ni, Zn, Co, Mn, and Cu, and nearly all of them were found to suppress $T_C$.[7,8] The suppression of $T_C$ is caused by the narrowing of the Ru $4dt_{2g}$ band induced by the change in Ru oxidation states and the destructive interaction between the Ru $t_{2g}$ bands via $e_g$ electrons of the replaced cations.[7] However, 10% Cr substitution for the Ru site of bulk SRO was reported to enhance $T_C$ up to 188 K.[7,9-12] A similar tendency was observed in thin films: SrRu$_{0.9}$Cr$_{0.1}$O$_3$ (SRCO) thin films epitaxially grown on the SrTiO$_3$(001) (STO) substrate have a higher $T_C$ (163 K) than SRO films (152 K); however, these values are lower than those of the corresponding bulk samples because of compressive strain from the STO substrate.[13,14] This implies that the influence of Cr substitution is independent of the epitaxial strain effect. The $T_C$ enhancement is attributable to the widening of the energy bands due to the hybridization of Cr $3dt_{2g}$ and Ru $4dt_{2g}$,[7] as predicted by the first principles band calculations.[15,16] The nuclear magnetic resonance of SrRu$_{1-x}$Cr$_x$O$_3$ ($x$ = 0.05 and 0.12) demonstrated that Cr exists as Cr$^{3+}$ ($t_{2g}^{3\uparrow}$), while Ru exists in a mixed valence state of Ru$^{4+}$ ($t_{2g}^{3\uparrow1\downarrow}$) and Ru$^{5+}$ ($t_{2g}^{3\uparrow}$) with less localized spin-down electrons in the Ru $4d$ shell.[10] This result is consistent with a widening of Ru $4dt_{2g}$ band and suggests a Ru$^{4+}$($d^4$)-O$^{2-}$-Ru$^{5+}$($d^3$) as well as Ru$^{4+}$($d^4$)-O$^{2-}$-Cr$^{3+}$($d^3$) double-exchange interaction.[10]

Another interesting feature of SRO thin films is their unusual behavior with regard to the anomalous Hall effect (AHE) at low temperature. In ferromagnetic metals, the Hall resistivity $\rho_H$ is generally described by a combination of the ordinary component $R_0\mu_0 H$, proportional to the external field $H$, and the anomalous component $R_s\mu_0 M$, proportional to the magnetization $M$, where $R_0$ and $R_s$ are the ordinary and anomalous Hall coefficients, respectively. Above 60 K, the $R_s$ value of the SRO thin film can be expressed by a combination of the skew and the side-jump terms as $R_s = a\rho + b\rho^2$, where the coefficients $a$ and $b$ reflect the characteristics of the asymmetric scatterings and $\rho$ is resistivity.[18] However, below 60 K, $R_s$ does not follow the above-mentioned relation, probably owing to a structural change at ~60 K with the modulation of the Fermi surface.[18] In contrast, $R_s$ of a $Ba_{0.2}Sr_{0.8}RuO_3$ film grown on the STO substrate does not show such an anomaly in the temperature range of 5–150 K; this is possibly because the structural change responsible for the anomaly is suppressed by the stress from the STO substrate, which exhibits a higher lattice mismatch with respect to $Ba_{0.2}Sr_{0.8}RuO_3$ than to SRO.[18] SRCO has almost the same lattice constants as SRO. Therefore, the AHE of the SRCO thin film would provide information about the effect of *B*-site substitution on structure.

In this study, we fabricated SRO and SRCO epitaxial thin films on STO substrates by pulsed laser deposition (PLD) and investigated their magnetic and transport properties as well as electronic states. As a result, resonant photoemission spectroscopy (PES) revealed that the Cr $3dt_{2g}$ orbital is hybridized with the Ru $4dt_{2g}$ orbital in the SRCO film. In addition, we found that the $R_s$ value of the SRCO thin film shows an anomaly in the ferromagnetic region, suggesting that the SRCO film undergoes a structural transition at low temperature with the modulation of the Fermi surface.

## II. EXPERIMENT

SRO and SRCO thin films were grown on STO substrates through PLD using a KrF excimer source with a laser fluence of 2.3 J/cm$^2$ per shot (5 Hz, wavelength = 248 nm). As PLD targets, we used SRO and SRCO ceramic pellets prepared from mixed powders of $SrCO_3$, $RuO_2$, and $Cr_2O_3$ by a solid-state reaction (pre-sintering at 1073 K for 12 h and sintering at 1373 K for 12 h). The substrate temperature and oxygen partial pressure were kept at 873 K and 13.3 Pa, respectively, during each deposition run. The typical thickness of the obtained films was ~50 nm.

Crystal structures of the SRO and SRCO films were examined by X-ray diffraction (XRD) using Cu-$K\alpha$ radiation. Magnetic properties were measured using a superconducting quantum interference device (SQUID) magnetometer. In the magnetization versus temperature ($M$–$T$) measurements, the films were first cooled to 5 K under $H$ = 4 MA/m, and then $M$ was measured by heating the sample under $H$ = 0.04 MA/m, applied perpendicular to the film surfaces. The in-plane electrical resistivities were measured using the four-probe method with indium electrodes. The Hall resistivities were characterized using a six-terminal Hall bar with a width of 1 mm and a length of 3 mm, where a magnetic field of 7 MA/m was applied perpendicular to the film surfaces. PES and X-ray absorption spectroscopy (XAS) measurements were performed at beamline 2C of Photon Factory, KEK. The binding energy was determined relative to the Fermi edge of a gold foil in electrical contact with the samples. The XAS spectra were measured using the total-electron-yield method.

### III. RESULTS AND DISCUSSION

Figure 1(a) shows out-of-plane $2\theta$-$\theta$ XRD patterns of the SRO and SRCO thin films, clearly indicating the 002 peaks of perovskite SRO and SRCO appear at a lower angle region compared with the 002 peak of STO. No secondary phase was detected. Figures 1(b) and (c) display the XRD reciprocal-space maps for asymmetric 103 diffractions of the SRO and SRCO thin films. As seen in the figures, the $q_x$ values of the 103 peaks coincide with that of the 103 peak of STO, proving that these films were coherently grown on STO substrates; that is, the SRO and SRCO films have the same in-plane lattice parameter as STO (0.3905 nm). With Cr substitution, the out-of-plane lattice parameter decreased from 0.3946 nm in the SRO film to 0.3938 nm in the SRCO film, reflecting that the ionic sizes of $Cr^{3+}$ (0.0615 nm) and $Ru^{5+}$ (0.0565 nm) are smaller than that of $Ru^{4+}$ (0.062 nm).[7]

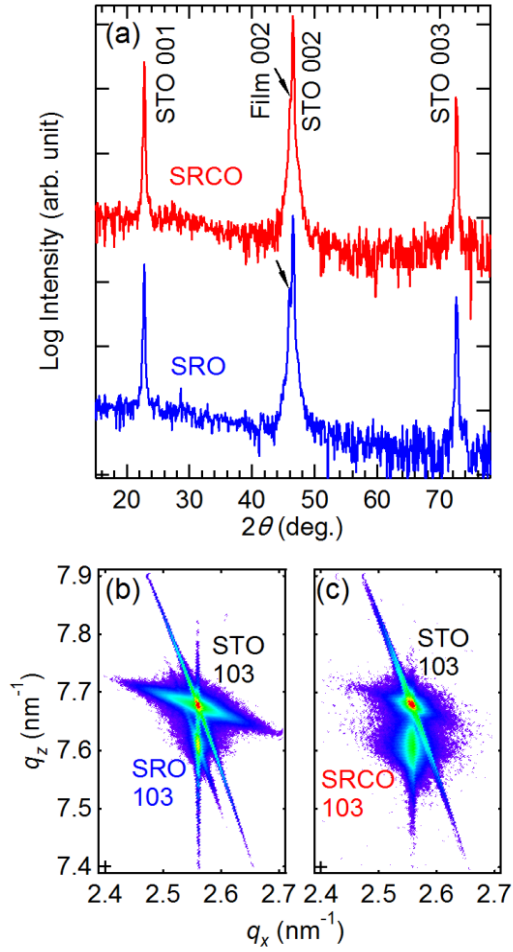

FIG. 1. (Color online) (a) Out-of-plane $2\theta$-$\theta$ XRD patterns for the SRO and SRCO thin films. The arrows indicate the 002 peaks of the SRO and SRCO films. XRD reciprocal-space maps corresponding to the asymmetric 103 diffraction of the (b) SRO and (c) SRCO thin films.

The unit-cell volumes calculated from these lattice parameters were $6.017 \times 10^{-2}$ nm$^3$ and $6.005 \times 10^{-2}$ nm$^3$ for the SRO and SRCO films, respectively; these values are slightly smaller than those of the corresponding bulk SRO and SRCO samples, which were $6.049 \times 10^{-2}$ nm$^3$ and $6.017 \times 10^{-2}$ nm$^3$,[9] respectively, because of the compressive strain from the STO substrate.

First, we checked the quality of the SRO and SRCO films by measuring their magnetic and transport properties. Figures 2(a) and (b) show in-plane and out-of-plane hysteresis $M$–$H$ curves for the SRO and SRCO films, respectively, measured at 5 K. For both films, the out-of-plane magnetization was two or three times larger than the in-plane one, indicating that the easy axes of these films are in the out-of-plane direction. However, the out-of-plane saturation magnetization $M_s$ of the SRO film was 1.8 $\mu_B$/f.u., which is larger than that of the SRCO film ($M_s$ = 1.1 $\mu_B$/f.u.), because of the enhanced antiferromagnetic coupling between Ru and Cr ions, as discussed in Ref. [9]. As seen from the out-of-plane magnetization vs temperature ($M$–$T$) plot in Fig. 2(c), the $T_C$ value of the SRCO film (166 K) is higher than that of the SRO film (147 K). These values are in good agreement with those reported previously.[17]

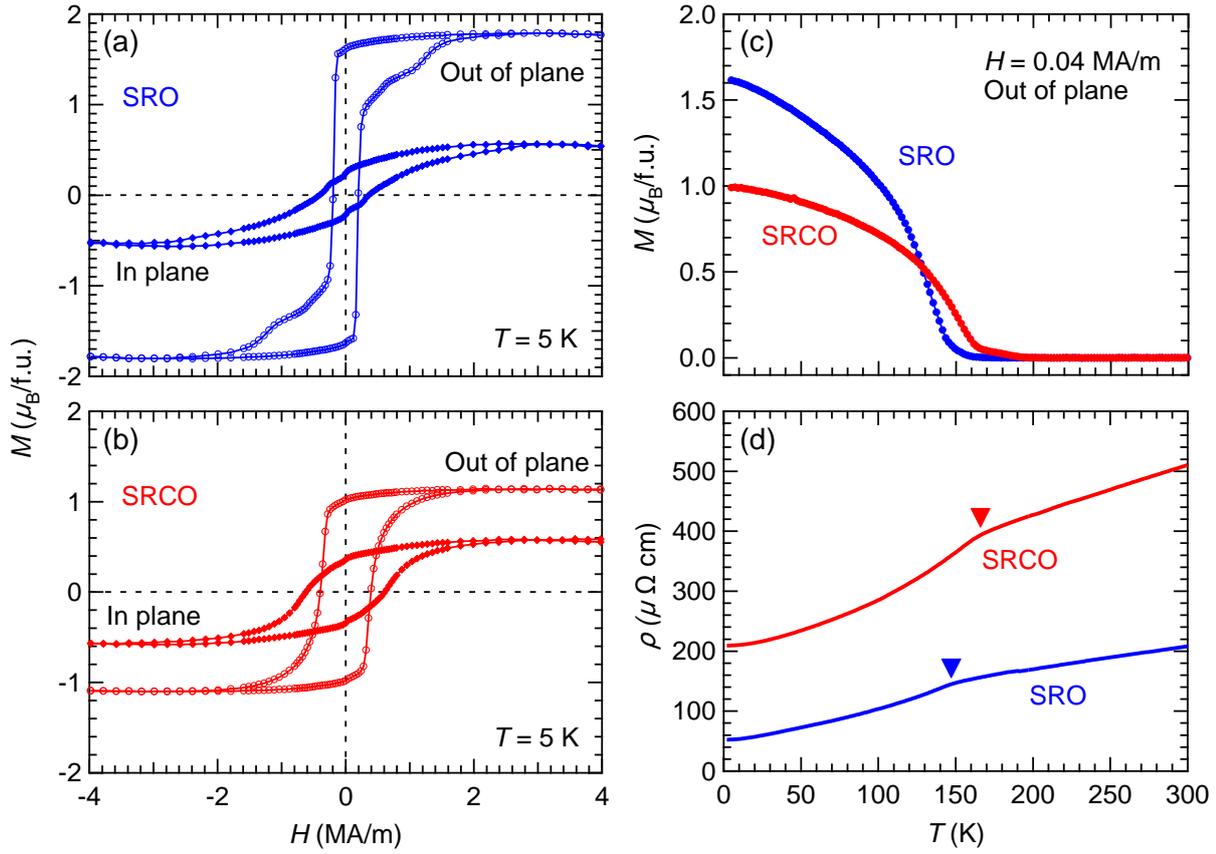

FIG. 2. (Color online) In-plane and out-of-plane hysteresis $M$–$H$ curves for the (a) SRO and (b) SRCO films measured at 5 K. (c) Out-of-plane $M$–$T$ curves for the SRO and SRCO films. (d) In-plane $\rho$–$T$ curves for the SRO and SRCO films. The filled triangles indicate $T_C$ values evaluated from the $M$–$T$ curves.

The in-plane resistivity as a function of temperature ($\rho$-$T$) for the SRO and SRCO films is shown in Fig. 2(d). Both films exhibited metallic $\rho$−$T$ behavior (d$\rho$/d$T$ > 0), and $\rho$ of the SRCO film was higher than that of the SRO film below 300 K. The residual resistivity ratio (RRR), which is defined as the ratio of the resistivity value at 300 K to that at 5 K, was 4.0 for SRO and 2.4 for SRCO, which are nearly the same as the values reported previously.[17] Both curves have kinks near the $T_C$ evaluated from the $M$–$T$ curves, which can be attributed to a Fisher–Langer anomaly associated with the rapid decrease in carrier scattering due to spin disorder below the transition temperature.[19]

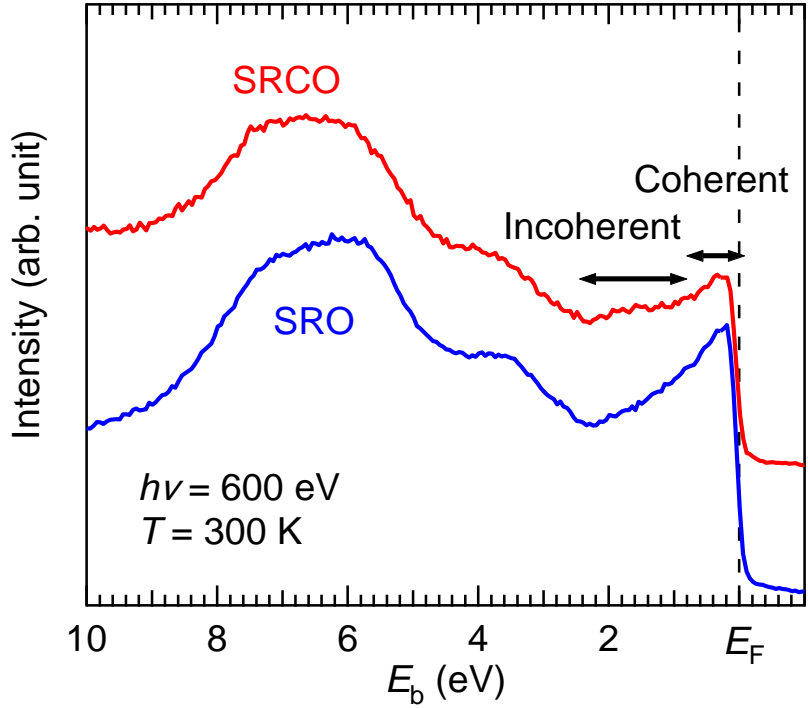

FIG. 3. (Color online) Valence-band PES spectra of the SRO and SRCO films.

Next, we elucidate the change in the density of states with Cr substitution from the valence-band PES spectra of the SRO and SRCO films, as shown in Fig. 3. Both films demonstrate finite spectral intensity at the Fermi energy ($E_F$), which is an indicator of metallicity. The valence band ranging from a binding energy ($E_b$) of ~2.5 eV to $E_F$ is mainly composed of Ru 4$d$ states and that at $E_b$ = 2.5–10 eV is assignable to O 2$p$ – Ru 4$d$ hybridized states.[6] The Ru 4$d$-derived bands split into two: a band around 1 eV with a sharp Fermi edge corresponding to the coherent part of the spectral function and a broad band centered at 2 eV corresponding to the incoherent part of the spectral function.[6] Comparison of the two spectra reveals that with Cr substitution, the spectral weight is transferred from the coherent part near $E_F$ to the incoherent part around $E_b = 2$ eV. This result implies that the increase in $T_C$ with Cr substitution is not associated with the Stoner instability via the enhanced density of states at $E_F$. A similar spectral weight transfer was observed in $Ca_{1-x}Sr_xRuO_3$ and has been explained by the increased correlation strength due to the lifting of the $t_{2g}$ orbital degeneracy.[6] Therefore, the present result suggests that the correlation strength of SRO is enhanced with Cr substitution.

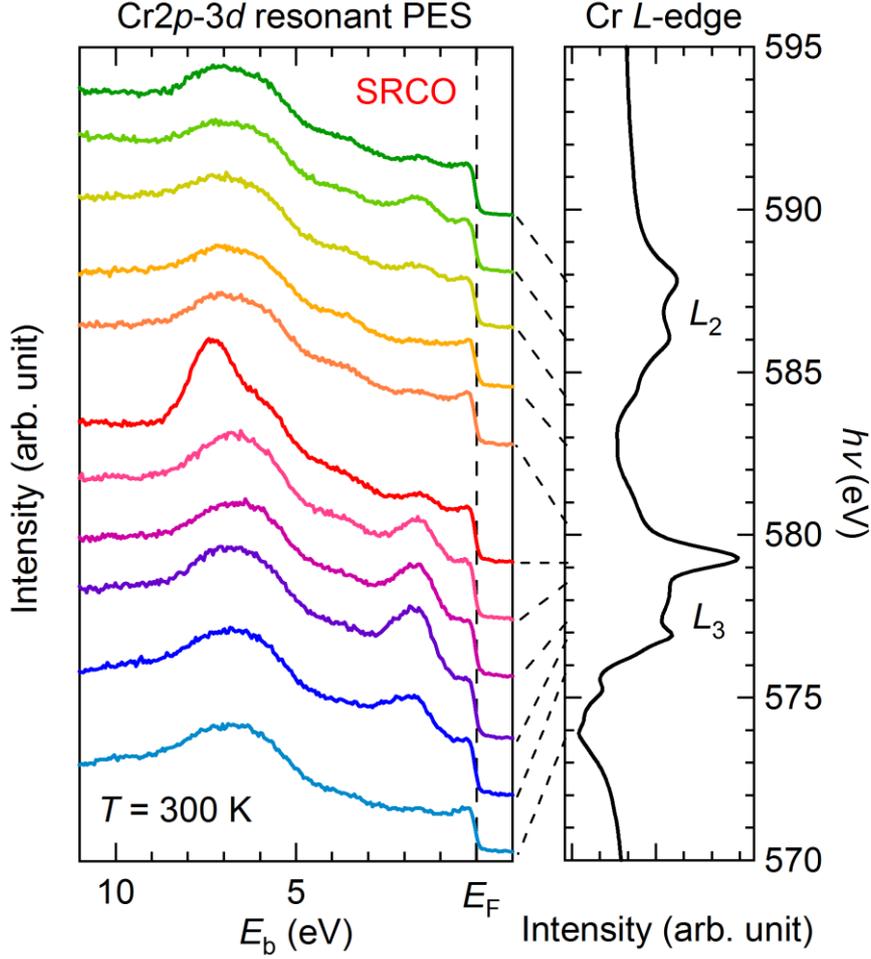

FIG. 4. (Color online) (Left) Cr 2p–3d resonant PES spectra of the SRCO film recorded at different excitation photon energies across the Cr 2p-3d transition. The broken lines indicate the photon energies (hv) at which the PES spectra were taken. (Right) Cr L-edge XAS spectrum of the SRCO film.

According to the first principles band calculations, the Ru $4dt_{2g}$ band of SRO widens with Cr substitution because of the hybridization of Cr $3dt_{2g}$ and Ru $4dt_{2g}$.[15,16] To investigate the contribution of Cr $3dt_{2g}$ states to the valence band of the SRCO film, the XAS spectrum near the Cr $L$-edge and the Cr $2p$-$3d$ resonant PES spectra were measured, as shown in Fig. 4. The resonant PES spectra were recorded over an excitation energy range of 573–588 eV and normalized by the photon flux of the incident beam. As shown in Fig. 4, the PES intensities at $E_b$ of 1–3 eV and 6–9 eV are resonantly enhanced, indicating that the corresponding energy states are composed of the Cr $3d$ orbital. We assigned the peaks located at 0–3 eV and 5–9 eV to Cr $3dt_{2g}$-derived states. By comparing the resonant PES spectra with the valence-band spectra of the SRO thin film, we found that the incoherent part of the Ru $4dt_{2g}$ states is located at the same energy levels as the Cr $3dt_{2g}$ states in the valence band. Therefore, the study of the resonant PES spectra provides experimental evidence that the Cr $3dt_{2g}$ orbital is hybridized with the Ru $4dt_{2g}$ orbital in the SRCO film, supporting the fact that the $T_C$ enhancement with Cr substitution stems from the widening of the energy bands due to the hybridization of these orbitals. The widened energy bands suggest the $Ru^{4+}$-$O^{2-}$-$Ru^{5+}(d^3)$ and $Ru^{4+}(d^4)$-$O^{2-}$-$Cr^{3+}(d^3)$ double-exchange interaction mentioned earlier.

As mentioned in the introduction section, the AHE of the SRO epitaxial film on an STO substrate exhibits an anomaly below 60 K, which possibly originates from the structural change with the modulation of the Fermi surface.[18] This structural change was observed using synchrotron XRD measurements at low temperatures under a magnetic field of ~0.16 MA/m perpendicular to the film,[18] although there is no signature of the structural change in the $\rho$–$T$ and $M$–$T$ curves. In order to investigate the influence of Cr substitution on AHE, we performed Hall effect measurements of the SRO and SRCO films. Figure 5 compares the magnetic-field dependences of $\rho_H$ ($\rho_H$–$H$) of the films measured at 5 K and 120 K. The $\rho_H$–$H$ curves of the SRO and SRCO films at 5 K showed clear hysteresis with coercivities ($H_c$) of 0.14 MA/m and 0.35 MA/m, respectively; these values are close to those obtained from the out-of-plane $M$−$H$ curves, as shown in Figs. 2(a) and 2(b). The $\rho_H$ value of the SRCO film at $H$ = 4 MA/m (1.26 $\mu\Omega$ cm) is much larger than that of the SRO film (0.31 $\mu\Omega$ cm), which may be due to the carrier scattering by substituted Cr ions. Notably, as shown in the inset of Fig. 5(a), it is clear that the $\rho_H$ of the SRO film at 5 K is not a linear function of $H$ even above 0.5 MA/m, at which the out-of-plane $M$ was saturated. A similar non-linear AHE with respect to $H$ has been reported previously.[18] In contrast, the $\rho_H$ value at 120 K did not show such non-linearity. A non-linear AHE was also observed in the 10% Cr-doped sample at 5 K, as shown in the inset of Fig. 5(b), though the magnitude of non-linearity in the SRCO film was much smaller than that in SRO.

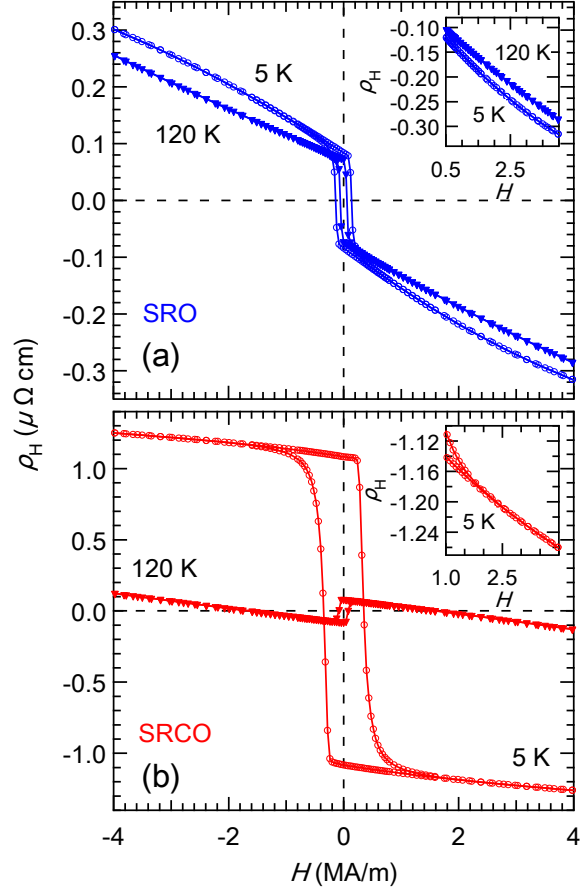

FIG. 5. (Color online) Magnetic-field dependence of Hall resistivity of the (a) SRO and (b) SRCO films measured at 5 K and 120 K. The insets show magnified views of the high-field regions for the SRO and SRCO films.

In order to investigate the AHE of the SRO and SRCO films in more detail, we analyzed the temperature dependences of $R_0$ and $R_s$, which were obtained by fitting the field dependence of $\rho_H$ using $\rho_H = R_0\mu_0 H + R_s\mu_0 M$, where $M$ is the measured out-of-plane magnetization. Figure 6 plots the temperature dependences of $R_0$ and $R_s$ for the SRO and SRCO thin films in the ferromagnetic region. In both the SRO and SRCO films, $R_0$ was nearly independent of temperature, while $R_s$ exhibited significant temperature dependence, as shown in Figs. 6(a) and 6(b). In particular, the $R_s$ value of the SRO film showed a minimum at ~80 K, as shown in Fig. 6(a). This nonmonotonic temperature dependence of $R_s$ can be attributed to the extrinsic side-jumps and the intrinsic Karplus–Luttinger mechanisms, which is associated with Fermi surface morphology, as reported by N. Haham *et al.*[20,21] On the other hand, $R_s$ of the SRCO film monotonically decreased with decreasing temperature (Fig. 6(b)). This suggests that the extrinsic effect associated with Cr substitution is dominant compared with the Karplus–Luttinger mechanism in the SRCO film. With increasing of the density of scattering centers (Cr), the extrinsic side-jumps contribution would become more prominent. In addition, the sign-change temperatures at which $R_s$ became zero were ~124 K and ~113 K in the SRO and SRCO films, respectively. The sign change of $R_s$ with temperature suggests that broad regions of the Fermi surface have nearly zero band curvature.[22]

To separate the intrinsic and extrinsic contributions to $R_s$, we plot $R_s/\rho$ as a function of $\rho$ ($R_s/\rho-\rho$) for the SRO and SRCO films, as shown in Fig. 7. The extrinsic part of $R_s$ is expressed as $a\rho + b\rho^2$, where the coefficients $a$ and $b$ are the parameters of skew and side-jump scattering, respectively, and $R_s/\rho$ shows a linear relation with $\rho$. As shown in the figures, linear $R_s/\rho-\rho$ relations were observed above 80 K for the SRO film (Fig. 7(a)) and above 60 K for the SRCO film (Fig. 7(b)). These linear relations prove that AHE can be explained by the skew and side-jump scatterings in these temperatures ranges. On the other hand, the $R_s/\rho-\rho$ plots deviated from the linear relations below certain temperatures: ~ 60 K for the SRO film and ~ 40 K for the SRCO film. These results indicate that the coefficients $a$ and $b$ are not constant in the entire ferromagnetic region of the SRO and SRCO films, suggesting that the modulation of the Fermi surface occurs at low temperature. This $R_s/\rho-\rho$ behavior of the SRO film is similar to that reported previously.[18]

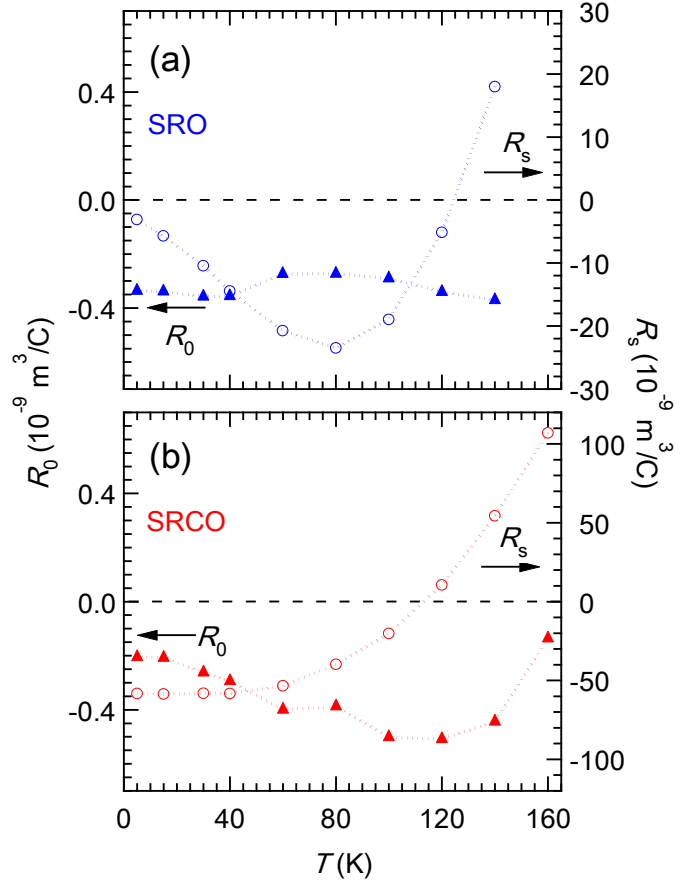

FIG. 6. (Color online) Temperature dependence of the ordinary and extraordinary Hall coefficients ($R_0$ and $R_s$, respectively) for the (a) SRO and (b) SRCO thin films in the ferromagnetic region.

In Fig. 7(b), we also observed non-linear relations in the $R_s/\rho-\rho$ plots of the SRCO films in the ferromagnetic region, although the magnitude of deviation from the fitting curve (green dashed line) is smaller than that of the SRO film. This suggests that the AHE of the SRCO thin film includes the intrinsic Karplus–Luttinger term, and that the SRCO thin film undergoes a structural transition at low temperature with the modulation of the Fermi surface. According to the previous report by Y. Kobayashi *et al.*, the stress from the STO substrate suppresses the structural phase transition of $Ba_{0.2}Sr_{0.8}RuO_3$ which has a higher lattice mismatch than SRO.[18] In this study, however, the structural transition in the SRCO thin film is not affected by the stress from the STO substrate, because the ionic radii of $Cr^{3+}$ (0.0615 nm) and $Ru^{4+}$ (0.062 nm) are nearly identical. Note that neither SRO nor SRCO showed distinctive features on the $\rho-T$ or $M-T$ curves at the temperatures at which kinks were observed in the $R_s/\rho-\rho$ plots. Specific experiments on the crystal and magnetic structures for SRCO are necessary to elucidate the effect of Cr substitution on AHE comprehensively.

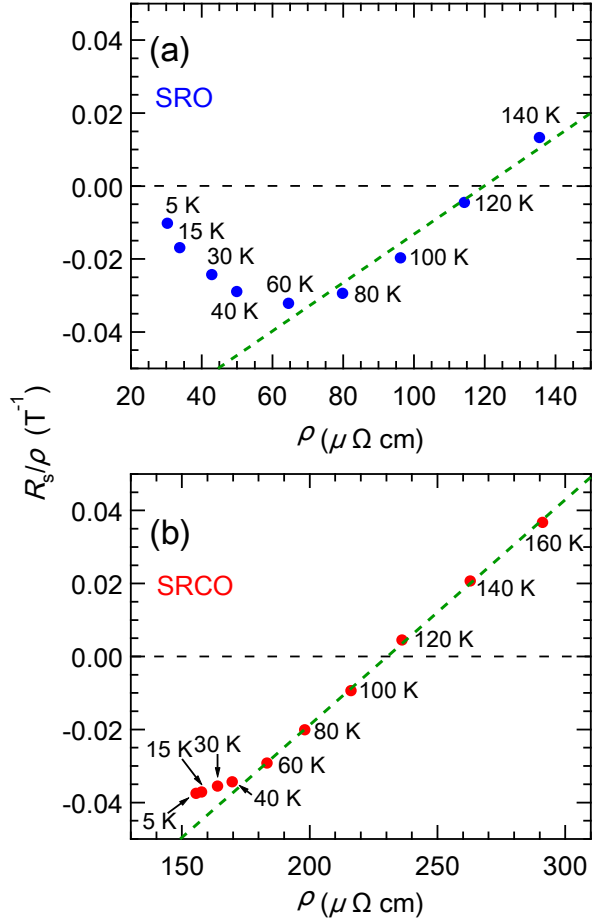

FIG. 7. (Color online) $R_s/\rho$ versus $\rho$ for the (a) SRO and (b) SRCO films. Linear fittings for the SRO and SRCO films (green dashed lines) were performed in the temperature ranges of 60−140 K and 60−160 K, respectively.

## IV. CONCLUSION

We investigated the magnetic and transport properties and the electronic states of SRO and SRCO thin films epitaxially grown on STO substrates. The $T_C$ of the SRCO film (166 K) was higher than that of the SRO film (147 K), while the SRCO film showed higher resistivity and lower RRR than the SRO film owing to the impurity scattering caused by Cr substitution. Cr 2p-3d resonant PES measurements revealed that the Cr $3dt_{2g}$ orbital is hybridized with the Ru $4dt_{2g}$ orbital in the SRCO film, supporting the assumption that the enhancement of the ferromagnetic transition temperature through Cr substitution stems from the widening of the energy bands due to the hybridization of Cr $3dt_{2g}$ and Ru $4dt_{2g}$ orbitals. Furthermore, the Hall resistivity of the SRCO film showed an anomaly in the ferromagnetic region, implying that the SRCO film undergoes a structural transition at low temperature with the modulation of the Fermi surface.

## ACKNOWLEDGMENTS

This work was partially supported by the Murata Science Foundation and JSPS KAKENHI Grant Number 25790054. Synchrotron radiation experiments were performed under the approval of the Photon Factory Program Advisory Committee, KEK (No. 2011G113 and 2011S2-003).